# Switchable high-$Q$ light absorbers based on phase-change resonant metasurfaces

Kai Qi[1,2], Guoxiang Wang[2], Xiang Shen[2], Yixiao Gao[1,2,*]

[1]School of Information and Control Engineering, China University of Mining and Technology, Xuzhou, 221116, Jiangsu, China

[2]Laboratory of Infrared Materials and Devices, Zhejiang Key Laboratory of Photoelectric Detection Materials and Devices, Research Institute of Advanced Technologies, Ningbo University, Ningbo, Zhejiang 315211, China

*gaoyixiao@nbu.edu.cn

*Abstract*: In this paper, we propose a switchable high-$Q$ light absorber based on a reconfigurable metasurface enabled by a lowloss phase-change material (PCM). By leveraging the coupling between guided-mode resonance and Fabry–Perot modes, mediated by the phase-transition dynamics of the embedded PCM, the resonance $Q$ factor can be actively tuned. This allows the system to switch from a perfect dark state, governed by the physics of bound states in the continuum, to a critically coupled resonance with a finite $Q$ factor. Consequently, the metasurface exhibits perfect absorption in the amorphous state and a reflection-dominated response in the crystalline state. The proposed metasurface holds significant potential for diverse nanophotonic applications, including photodetection and thermal emission control.

## 1. Introduction

Light absorption in thin films is central to a wide range of optoelectronic applications. Broadband absorption is essential for solar energy harvesting [1] and photodetection [2], whereas narrowband absorption is crucial for thermal emission control [3] and sensing [4,5], where the realization of high-$Q$ resonances in flat optics is particularly important [6]. High-$Q$ resonances can be realized through guided-mode resonances in grating structures [7] and surface lattice resonances in metallic particle arrays [8]; however, achieving higher $Q$ factors often demands reduced unit-cell feature sizes, which introduces significant nanofabrication challenges. Recently, resonant metasurfaces governed by the physics of bound states in the continuum (BICs) have emerged as a have emerged as a powerful strategy for realizing high-$Q$ resonances [9]. An ideal BIC represents a dark state inaccessible to far-field excitations. By introducing symmetry breaking [9,10] or fine-tuning geometric parameters [11–13], this dark state can be converted into a quasi-BIC (q-BIC) with a controllable radiation rate. Such q-BIC resonances in metasurfaces have been widely exploited for lasing [14], optical nonlinearity [15,16], sensing [17], etc.

High-$Q$ perfect light absorption relies on satisfying the critical coupling condition [18,19], in which the radiative and nonradiative mode decay rates are balanced to maximize absorptance at the resonance. The tunable $Q$ factor of q-BICs in metasurfaces provides an excellent platform for achieving perfect absorption via critical coupling. Symmetry-broken metasurfaces with tailored $Q$ factors have been employed to enhance light absorption in two-dimensional materials [20], weakly absorbing media (e.g. germanium nanostructures) [21,22], etc. To mitigate the polarization dependence introduced by symmetry breaking, metasurfaces with multi-element unit cells have been designed, and by perturbing the $C_{4v}$ symmetry of the unit cell, polarization-independent perfect absorbers can be realized [23,24]. In addition, the concept of merging BICs has emerged as an effective strategy for constructing high-$Q$ resonances with polarization insensitivity [11,12,25], where the $Q$ factor can be precisely tuned through unit-cell parameter adjustments. Recently, Qi *et al.* demonstrated a polarization-independent perfect absorber design that leverages both radiative and nonradiative loss engineering : the radiative loss is tuned by adjusting the lattice period to merge the Γ-point radiative guided-mode resonance with dark accidental BICs, while the nonradiative loss is controlled by varying the separation between an embedded waveguiding layer and the absorbing medium [26]. However, in these approaches, the absorption behavior remains fixed once the devices are fabricated.

Active tuning of perfect absorption has therefore become an important research direction, with the incorporation of tunable optical materials into metasurfaces offering a promising approach. Materials such as graphene [27], epsilon-near-zero (ENZ) media [28], have been explored to realize perfect absorbers with spectrally tunable responses. Among them, chalcogenide phase-change materials (PCMs) are particularly attractive due to their large refractive index contrast and nonvolatile phase transitions triggered by electrical or optical stimuli, making them ideal candidates for reconfigurable photonic devices [29]. Tunable perfect absorbers have been demonstrated using $Ge_2Sb_2Te_5$ embedded in multilayer structures [30,31] and metallic metasurfaces [32,33], where phase transitions induce pronounced spectral shifts in absorption.

In this work, we propose a switchable light absorber based on a low-loss PCM–enabled reconfigurable resonant metasurface, which can toggle between a high-$Q$ perfect absorption state and a near-total reflection state through phase transitions. We first present the operating principle of the proposed absorber using temporal coupled-mode theory. We then investigate $Q$-factor control arising from the coupling between guided-mode resonance (GMR) and a background Fabry–Perot (FP) mode in the metasurface. Building on this mechanism, we demonstrate a switchable absorber enabled by low-loss PCM-mediated GMR–FP coupling. Finally, we analyze the impact of fabrication imperfections on device performance.

## 2. Principle and design

Figure 1(a) illustrates a BIC-driven reflective metasurface composed of a lossy dielectric disk array positioned above a mirror. Owing to the presence of the back reflector, the structure behaves as a one-port resonant system that supports a single mode, and can be accurately described using temporal coupled-mode theory (TCMT) [34].

$$\frac{da}{dt} = \left(i\omega_0 - \gamma_r - \gamma_n\right)a + \kappa s_+ \quad (1)$$
$$s_- = s_+ - \kappa a$$

where $a$ is the resonant-mode amplitude with an angular frequency of $\omega_0$, and $\gamma_r$ and $\gamma_n$ denote the radiative and non-radiative decay rates, $\kappa$ is the coupling coefficient between resonance and incident wave, which is $\kappa = \sqrt{2\gamma_r}$ due to time reversal symmetry. and $s_+$ and $s_-$ represent the amplitudes of the incoming and outgoing waves. Under steady-state excitation $s_+(t) = s_+\exp(-i\omega t)$ and $a(t) = a\exp(-i\omega t)$. The reflection coefficient is $r(\omega) = s_-/s_+$, and the absorption spectrum $A = 1- |r(\omega)|^2$, which becomes

$$A(\omega) = \frac{4\gamma_r \gamma_n}{(\omega - \omega_0)^2 + (\gamma_r + \gamma_n)^2} \tag{2}$$

In the resonant systems, the resonance behavior is usually characterized by the quality factors, and the corresponding radiative and nonradiative $Q$ factor can be defined as $Q_r = \omega_0/2\gamma_r$ and $Q_n = \omega_0/2\gamma_n$.

A key advantage of BIC-driven resonances is their tunable radiative loss, which enables precise control of the absorption behavior. Two important operating regimes can be identified:

(1) Critical-coupling regime: when $\gamma_r = \gamma_n$, (equivalently $Q_r = Q_n$), the absorption at the resonant frequency reaches unity, corresponding to perfect absorption. Achieving a high-$Q$ perfect absorber therefore requires maximizing $Q_n$ [26].

(2) BIC (dark-state) regime: when $\gamma_r$ approaching zero, i.e. $Q_r$ is diverging, the external coupling coefficient $\kappa$ vanishes and the resonance becomes a perfect dark state that cannot be excited by the incoming wave. In this case the absorptance drops to zero, and the structure behaves as a near-perfect reflector.

In the following, we demonstrate how the proposed metasurface can be switched between the critical-coupling regime and the dark-BIC regime through the phase transition of the embedded PCM layer, as depicted in Fig. 1(b).

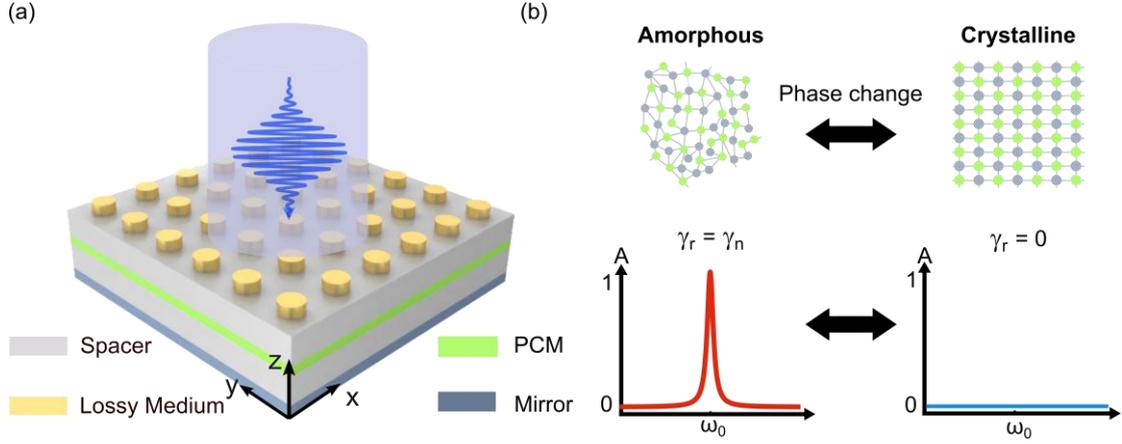

**Fig. 1.** (a) Schematic of the proposed switchable light absorber based on a phase-change metasurface. (b) Operating principle of the device, where the phase transition of the embedded PCM layer enables the metasurface to switch between a narrowband perfect-absorption state and a reflection-dominated state, $\gamma_r$ and $\gamma_n$ are the radiative and nonradiative loss rate of the studied resonant mode.

The underlying mechanism enabling such switchability originates from the interaction between GMR and the F–P background mode. Their coupling provides an effective route to realizing high-$Q$ resonances [35], and is closely related to the merging-BIC effect [11,12,25]. Figure 2(b) shows the electric field profile of the resonance of interest in a single unit cell, which is a doubly-degenerate guided mode resonance. Here, the mirror is treated as a perfect electric conductor (PEC). The metasurface is composed of germanium disks separated from the mirror by a SiO$_2$ layer of thickness $S$, and we set the parameters as $H = 170$nm, $D = 400$nm, $S = 480$nm and $P = 1145$nm. Figure 2(c) shows the radiative ($Q_r$) and nonradiative ($Q_n$) $Q$ factors as a function of $S$. It is observed that $Q_r$ exhibits a periodic diverging behavior with increasing $S$. This occurs because the guided-mode resonance destructively interferes with the background F-P resonances of different orders, leading to the suppression of radiative loss at multiple $S$ values [12]. Meanwhile, $Q_n$ shows an increasing trend with $S$, as the thicker SiO$_2$ spacer reduces the field overlap with the lossy germanium disks. Near each $Q_r$ diverging point, $Q_r$ and $Q_n$ intersect twice, indicating the formation of the critical coupling condition. A similar behavior is observed when tuning the period of the germanium disk array, as shown in Fig. 2(d), demonstrating that $Q_r$ can be controlled by adjusting both $S$ and $P$. This periodic $Q_r$ diverging behavior suggests that it is possible to switch $Q_r$ from matching $Q_n$ to effectively infinite, forming the basis for a switchable high-$Q$ absorber.

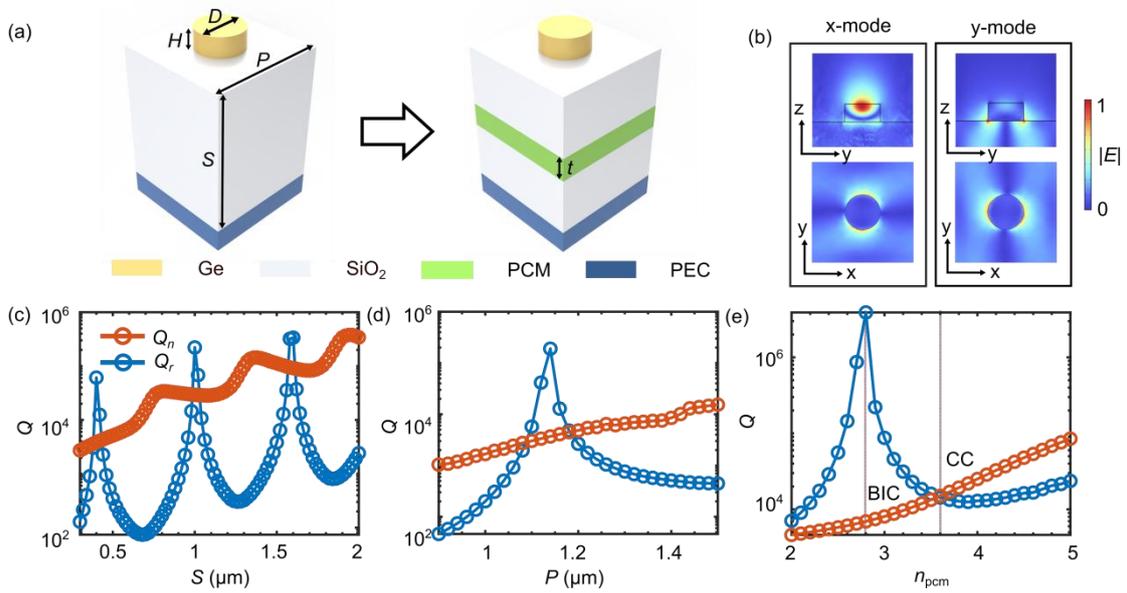

**Fig. 2.** (a) Schematic of the unit cells of the one-port resonant metasurface with and without the embedded PCM layer. (b) Electric field distribution of the degenerate resonant mode of interest. Variation of $Q_r$ and $Q_n$ as a function of (c) dielectric spacer thickness $S$, (d) metasurface period $P$ and (e) refractive index of the embedded PCM thin film $n_{PCM}$.

Actively engineering $Q_r$ through tuning the dielectric spacer thickness $S$ or metasurface period $P$ is technological challenging. However, the resonance behavior of the F-P mode depends on the optical path length, i.e. $n_sS$, where $n_s$ is the refractive index of dielectric spacer (e.g., SiO$_2$ in Fig. 2(a), indicating that F-P resonances can alternatively be controlled by tuning $n_s$. Considering that the dielectric spacer is typically a thin film, a material with a large refractive index change is preferred to realize a switchable light absorber. Chalcogenide PCMs have large refractive index contrast, for example, the classical Ge$_2$Sb$_2$Te$_5$ have a refractive index contrast of 1.1942 at 1550 nm. However, its large material absorption in the crystalline state has a negative impact on the high-$Q$ light absorption. Recently, the two-element PCM, Sb$_2$Se$_3$, present an ultralow loss phase change medium with a moderate refractive index contrast of 0.764 [36]. Here, we consider embedding a thin, lowloss PCM Sb$_2$Se$_3$ of thickness $t$ within the dielectric spacer, while keeping the total spacer thickness fixed at $S$, as depicted in Fig. 2(a). Figure 2(e) shows the variation of $Q_r$ and $Q_n$ with changes in the refractive index of the PCM $n_{PCM}$. In this example, $t = 25$ nm, $S = 425$ nm, and $P = 1145$ nm. It can be seen that when $n_{PCM}=2.8$, the resonance transforms into a BIC state, while increasing $n_{PCM}$ to 3.6 brings $Q_r$ into matching with $Q_n$, leading to perfect light absorption.

To sum up, the absorption characteristics of the proposed metasurface are primarily governed by the radiative quality factor $Q_r$. Embedding a thin PCM layer within the dielectric spacer allows the phase-transition–induced refractive-index change to modify the coupling between the GMR and F–P modes, thereby tuning $Q_r$ from matching $Q_n$ (critical coupling) to diverging toward infinity (dark BIC). This mechanism enables reversible switching between perfect absorption and high-reflectivity states.

## 3. Simulation and analysis

We next validate the performance of the PCM-enabled switchable light absorber through rigorous coupled wave analysis (RCWA) [37]. The germanium nanodisks are assumed to have a height $H = 150$ nm and diameter $D = 400$ nm. A $Sb_2Se_3$ thin film with thickness $t = 9$ nm is embedded in the middle of the dielectric spacer, whose total thickness is maintained at 359 nm. For simplicity, the mirror is treated as a perfect electric conductor (PEC).

Figure 3(b) shows the absorption spectra for varying metasurface period $P$ when $Sb_2Se_3$ is in its amorphous state. A narrowband absorption peak shifts toward longer wavelengths as $P$ increases. Two critical points are noteworthy: at $P = 1099$ nm, the peak absorptance reaches unity, indicating the formation of a perfect absorber, whereas at $P = 1048$ nm, the absorption peak vanishes owing to the formation of BIC resonance [12]. When the $Sb_2Se_3$ crystalized, a BIC is formed at $P = 1099$ nm, switching the perfect absorption state into a reflection-dominated state, with the peak absorptance decreasing by more than 20 dB. Figure 3(d) shows the absorption spectra at $P = 1099$ nm for the amorphous and crystalline phase of $Sb_2Se_3$, demonstrating the phase-change-induced switching of light absorption.

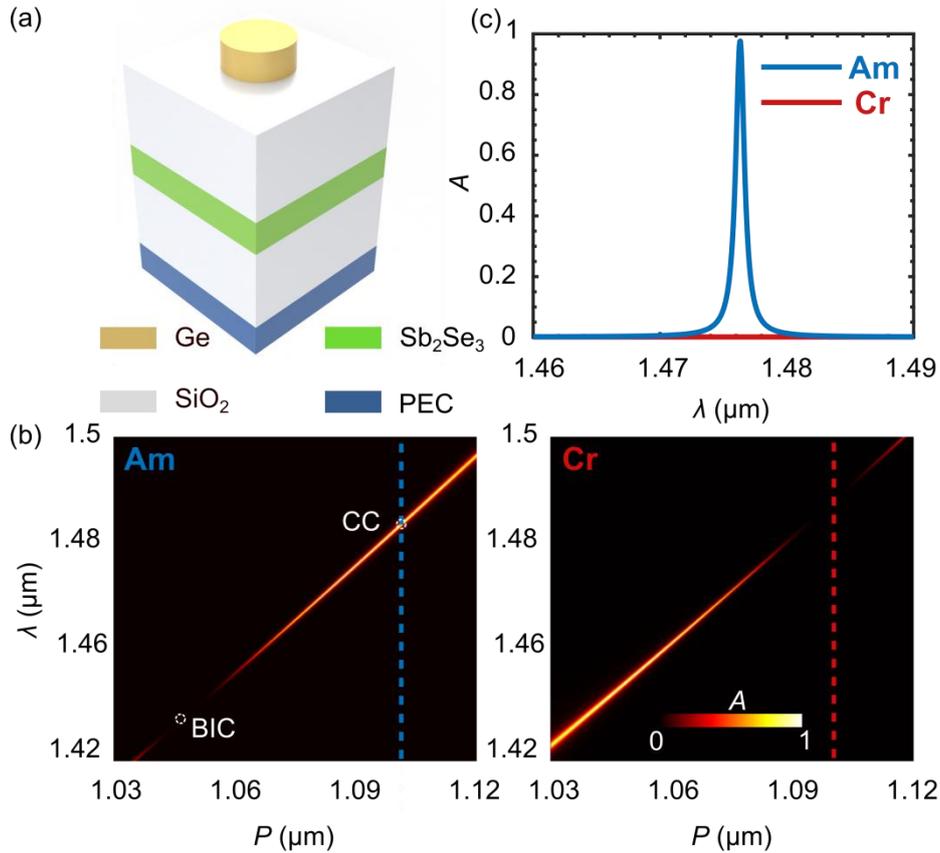

**Fig. 3.** (a) Schematic of the unit cell of the proposed resonant metasurface incorporating low-loss $Sb_2Se_3$. (b) Absorption spectra with varying metasurface period $P$ for the amorphous and crystalline states. (c) Absorption spectrum at $P = 1099$ nm for both phases of Sb2Se3.

In practice, low-loss noble metals such as gold and silver can serve as mirrors. To demonstrate realistic implementation, we replace the PEC substrate with a silver reflector, taking into account the material dispersion of both germanium (Ge) and silver (Ag) [38]. Figure 4(a) and 4(b) shows the absorption spectra for varying metasurface period $P$ with $Sb_2Se_3$ being amorphous and crystalline states, respectively. The metasurface structural parameters are set as $P = 1023$ nm, $H = 200$ nm and $S = 236$ nm. We could also observe that vanishing absorption linewidth at $P = 1023$ nm for amorphous $Sb_2Se_3$, and the BIC position would shift to longer wavelengths at larger metasurface period when $Sb_2Se_3$ is transformed to crystalline state. Figure 4(c) show the absorption spectra at $P = 1023$ nm for $Sb_2Se_3$ in amorphous and crystalline states. The linewidth of the absorption spectra is 8.5 nm, corresponding to a $Q$ factor of 180. Upon the phase change of $Sb_2Se_3$, an efficient absorptance modulation of 28 dB is achieved, switching the metasurface from a perfect absorber to a reflective state, confirming robust switching performance in this silver-mirror-based configuration.

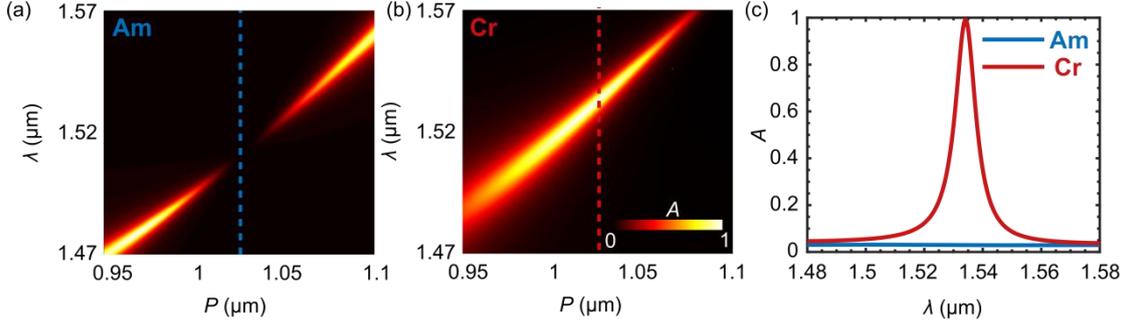

**Fig. 4.** Absorption spectra as a function of metasurface period $P$ for a silver-backed metasurface with $Sb_2Se_3$ in (a) amorphous and (b) crystalline phases. (c) Comparison of absorption spectra at $P = 1023$ nm for amorphous and crystalline $Sb_2Se_3$.

Fabricating the proposed light absorber requires multiple nanofabrication processes, including film deposition, pattern transfer, and etching. The resonance behavior is jointly influenced by the geometric parameters of the germanium nanodisk array and the film thicknesses, while the switching performance is particularly sensitive to variations in the film thickness. Here, we analyze the impact of dielectric spacer thickness variations on the performance of the switchable light absorber. Two major fabrication imperfections may occur during the film deposition process: (i) deviation of the PCM layer from the middle plane of the spacer, and (ii) thickness variation of the PCM layer itself. As illustrated in Fig. 5(a), we introduce a parameter $\Delta Z$ to characterize the displacement of the PCM layer from the spacer's mid-plane. $\Delta Z = 0$ corresponds to the PCM layer being precisely centered within the $SiO_2$ layers. $\Delta Z < 0$ indicates a thicker lower $SiO_2$ layer relative to the upper one, while $\Delta Z > 0$ represents the opposite scenario. Figure 5(b) shows the absorption spectra variation for different $\Delta Z$ values. For $\Delta Z = \pm 18$ nm, the originally designed BIC state for amorphous $Sb_2Se_3$ opens weak radiation channels, but the peak absorptance remains below 0.2. Meanwhile, the critical coupling state in crystalline $Sb_2Se_3$ experiences a slight reduction in absorptance, but maintains a minimum absorption exceeding 0.85 for $\Delta Z$ within $\pm 18$ nm.

We further examine the effect of thickness variations in the $Sb_2Se_3$ layer by introducing $\Delta t$, defined as the deviation from the designed thickness, as depicted in Fig. 5(c). Figure 5(d) shows the corresponding absorption spectra for $\Delta t$ within $\pm 3$ nm. The spectra for amorphous $Sb_2Se_3$ state show more noticeable changes, reaching an absorption of 0.25 at $\Delta t = +3$ nm, while the peak absorptance for the crystalline state remains above 0.9. Thickness variations in the $Sb_2Se_3$ layer also induce significant shifts in the resonance wavelength, with thicker layers causing a redshift. It should be noted that the peak absorptance contrast between the two phase states of $Sb_2Se_3$ is degraded by dielectric spacer thickness variations. However, this degradation can be mitigated by adjusting the lattice constant of the germanium nanodisk array, thereby restoring effective switching between the perfect absorption and reflection-dominated states. It is worth noting that, although the absorption characteristics are sensitive to variations in the film parameters, modern thin-film deposition technologies can reliably achieve nanoscale precision in thickness control. For example, Ref. [39] reports phase-change multilayers with reproducible thickness control at the 3-nm level, demonstrating that the accuracy required in our design is well within the capabilities of current deposition processes.

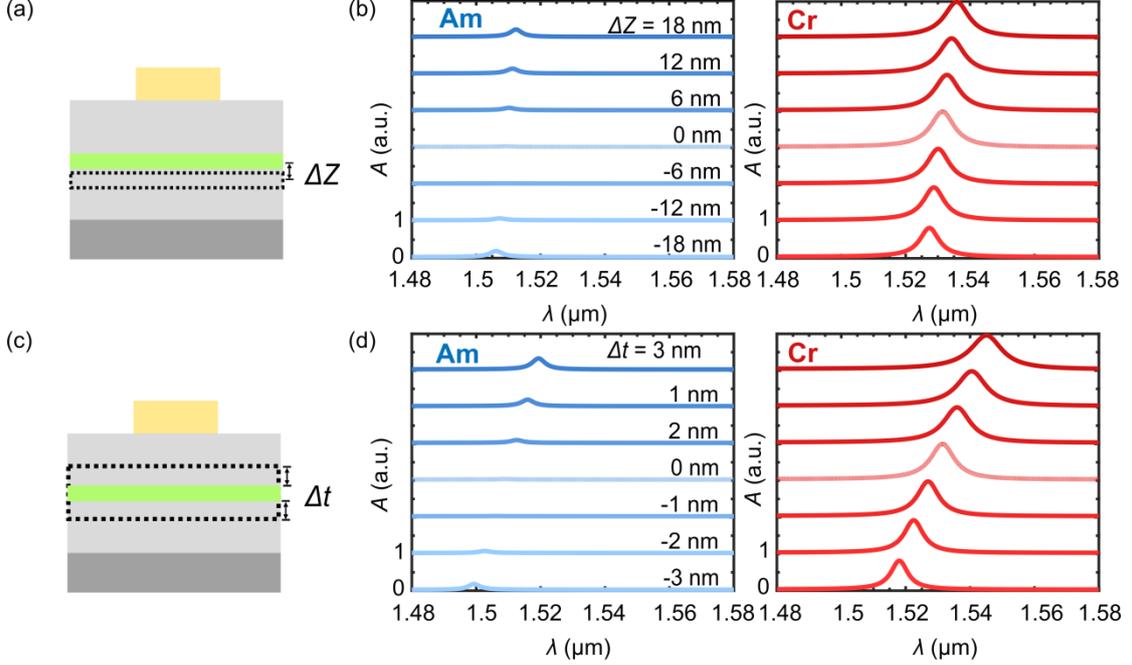

Fig. 5. (a) Schematic of the PCM layer deviating from the center of the SiO₂ spacer, characterized by $\Delta Z$. (b) Absorption spectra for different $\Delta Z$ values. (c) Schematic illustrating variations in PCM thickness, characterized by $\Delta t$. (b) Absorption spectra for different $\Delta t$ values.

For the potential application on the proposed devices, narrowband absorbers are highly desirable for optical sensing and photodetection [40,41], where spectral selectivity is essential for improving signal-to-noise ratio, suppressing background illumination, and enabling wavelength-specific detection. A switchable absorber/reflector further allows a single device to operate in multiple spectral-response states, thereby supporting reconfigurable sensing architectures without the need for additional optical components. In addition, the proposed design strategy can be readily extended to the mid-infrared spectral region, where narrowband absorbers serve as selective thermal emitters in accordance with Kirchhoff's law of thermal radiation. In this regime, a structure capable of toggling between a high-reflectivity state and a perfect-absorption state offers large-contrast emissivity modulation, enabling applications in dynamic thermal emission control [44], infrared camouflage [42], and adaptive radiative cooling [43].

## 4. Conclusion

In summary, we propose a switchable metasurface-based light absorber enabled by a low-loss phase-change material, capable of toggling between a high-$Q$ perfect absorption state and a reflection-dominated state. The switching functionality is achieved by controlling the coupling between guided-mode resonance and a background Fabry–Perot mode via the phase change, allowing the resonance to be either critically coupled or a perfect dark state. A practical metasurface design incorporating a silver back mirror demonstrates a perfect absorption state with a $Q$ factor of 180, which can be switched to a reflective state with over 28 dB change in absorptance. We also examine the impact of fabrication imperfections, showing that variations in the PCM layer thickness significantly affect performance, highlighting the need for precise thickness control during deposition. The proposed switchable metasurface offers promising potential for applications in narrowband photodetection, sensing, and thermal emission control.

**Acknowledgments**

**References**

1. L. Mascaretti, Y. Chen, O. Henrotte, O. Yesilyurt, V. M. Shalaev, A. Naldoni, and A. Boltasseva, "Designing Metasurfaces for Efficient Solar Energy Conversion," ACS Photonics **10**, 4079–4103 (2023).
2. J. Y. Suen, K. Fan, J. Montoya, C. Bingham, V. Stenger, S. Sriram, and W. J. Padilla, "Multifunctional metamaterial pyroelectric infrared detectors," Optica, OPTICA **4**, 276–279 (2017).
3. H. T. Miyazaki, T. Kasaya, M. Iwanaga, B. Choi, Y. Sugimoto, and K. Sakoda, "Dual-band infrared metasurface thermal emitter for CO2 sensing," Appl. Phys. Lett. **105**, 121107 (2014).
4. A. Tittl, A. Leitis, M. Liu, F. Yesilkoy, D. Y. Choi, D. N. Neshev, Y. S. Kivshar, and H. Altug, "Imaging-based molecular barcoding with pixelated dielectric metasurfaces," Science **360**, 1105–1109 (2018).
5. S. Luo, J. Zhao, D. Zuo, and X. Wang, "Perfect narrow band absorber for sensing applications," Opt. Express, OE **24**, 9288–9294 (2016).
6. J. Yu, W. Yao, M. Qiu, and Q. Li, "Free-space high-Q nanophotonics," Light Sci Appl **14**, 174 (2025).
7. S. S. Wang and R. Magnusson, "Theory and applications of guided-mode resonance filters," Appl. Opt., AO **32**, 2606–2613 (1993).
8. V. G. Kravets, A. V. Kabashin, W. L. Barnes, and A. N. Grigorenko, "Plasmonic Surface Lattice Resonances: A Review of Properties and Applications," Chem. Rev. **118**, 5912–5951 (2018).
9. K. Koshelev, S. Lepeshov, M. Liu, A. Bogdanov, and Y. Kivshar, "Asymmetric Metasurfaces with High-Q Resonances Governed by Bound States in the Continuum," Phys. Rev. Lett. **121**, 193903 (2018).
10. Y. Gao, J. Ge, S. Sun, and X. Shen, "Dark modes governed by translational-symmetry-protected bound states in the continuum in symmetric dimer lattices," Results Phys. **43**, 106078 (2022).
11. J. Jin, X. Yin, L. Ni, M. Soljacic, B. Zhen, and C. Peng, "Topologically enabled ultrahigh-Q guided resonances robust to out-of-plane scattering," Nature **574**, 501–504 (2019).
12. Y. Gao, J. Ge, Z. Gu, L. Xu, X. Shen, and L. Huang, "Degenerate merging BICs in resonant metasurfaces," Opt. Lett. **49**, 6633 (2024).
13. Y. Gao, L. Xu, and X. Shen, "Q-factor mediated quasi-BIC resonances coupling in asymmetric dimer lattices," Opt. Express **30**, 46680–46692 (2022).
14. Y. Zeng, X. Sha, C. Zhang, Y. Zhang, H. Deng, H. Lu, G. Qu, S. Xiao, S. Yu, Y. Kivshar, and Q. Song, "Metalasers with arbitrarily shaped wavefront," Nature **643**, 1240–1245 (2025).
15. J. T. Wang, P. Tonkaev, K. Koshelev, F. Lai, S. Kruk, Q. Song, Y. Kivshar, and N. C. Panoiu, "Resonantly enhanced second- and third-harmonic generation in dielectric nonlinear metasurfaces,"

Opto-Electron. Adv. **7**, 230186–15 (2024).

16. P. Hong, L. Xu, and M. Rahmani, "Dual bound states in the continuum enhanced second harmonic generation with transition metal dichalcogenides monolayer," Opto-Electron. Adv. **5**, 200097–8 (2022).

17. F. Yesilkoy, E. R. Arvelo, Y. Jahani, M. Liu, A. Tittl, V. Cevher, Y. Kivshar, and H. Altug, "Ultrasensitive hyperspectral imaging and biodetection enabled by dielectric metasurfaces," Nat. Photon. **13**, 390–396 (2019).

18. Y. Liu, A. Chadha, D. Zhao, J. R. Piper, Y. Jia, Y. Shuai, L. Menon, H. Yang, Z. Ma, S. Fan, F. Xia, and W. Zhou, "Approaching total absorption at near infrared in a large area monolayer graphene by critical coupling," Appl. Phys. Lett. **105**, (2014).

19. J. R. Piper, V. Liu, and S. Fan, "Total absorption by degenerate critical coupling," Appl. Phys. Lett. **104**, 251110 (2014).

20. X. Wang, J. Duan, W. Chen, C. Zhou, T. Liu, and S. Xiao, "Controlling light absorption of graphene at critical coupling through magnetic dipole quasi-bound states in the continuum resonance," Phys. Rev. B **102**, 155432 (2020).

21. J. Tian, Q. Li, P. A. Belov, R. K. Sinha, W. Qian, and M. Qiu, "High-Q All-Dielectric Metasurface: Super and Suppressed Optical Absorption," ACS Photon. **7**, 1436–1443 (2020).

22. J. Ge, Y. Gao, L. Xu, N. Zhou, and X. Shen, "Dual-symmetry-perturbed all-dielectric resonant metasurfaces for high-Q perfect light absorption," Chin. Opt. Lett. **22**, 023602 (2024).

23. R. Masoudian Saadabad, L. Huang, and A. E. Miroshnichenko, "Polarization-independent perfect absorber enabled by quasibound states in the continuum," Phys. Rev. B **104**, 235405 (2021).

24. K. Wu, H. Zhang, and G. P. Wang, "Polarization-independent high-Q monolayer Ge-assisted near-perfect absorber through quasibound states in the continuum," Phys. Rev. B **108**, 085412 (2023).

25. M. Kang, L. Mao, S. Zhang, M. Xiao, H. Xu, and C. T. Chan, "Merging bound states in the continuum by harnessing higher-order topological charges," Light Sci. Appl. **11**, 228 (2022).

26. K. Qi, J. Ge, X. Shen, and Y. Gao, "High-Q perfect light absorption enabled by degenerate merging BICs," Opt. Express **33**, 29869–29879 (2025).

27. Y. Yao, R. Shankar, M. A. Kats, Y. Song, J. Kong, M. Loncar, and F. Capasso, "Electrically Tunable Metasurface Perfect Absorbers for Ultrathin Mid-Infrared Optical Modulators," Nano Lett. **14**, 6526–6532 (2014).

28. S. S. Mirshafieyan and D. A. Gregory, "Electrically tunable perfect light absorbers as color filters and modulators," Sci Rep **8**, 2635 (2018).

29. M. Wuttig, H. Bhaskaran, and T. Taubner, "Phase-change materials for non-volatile photonic applications," Nat. Photon. **11**, 465–476 (2017).

30. V. K. Mkhitaryan, D. S. Ghosh, M. Rudé, J. Canet-Ferrer, R. A. Maniyara, K. K. Gopalan, and V. Pruneri, "Tunable Complete Optical Absorption in Multilayer Structures Including Ge2Sb2Te5 without Lithographic Patterns," Advanced Optical Materials **5**, 1600452 (2017).

31. K. V. Sreekanth, S. Han, and R. Singh, "Ge2Sb2Te5-Based Tunable Perfect Absorber Cavity with Phase Singularity at Visible Frequencies," Advanced Materials **30**, 1706696 (2018).

32. N. Mou, X. Liu, T. Wei, H. Dong, Q. He, L. Zhou, Y. Zhang, L. Zhang, and S. Sun, "Large-scale, low-cost, broadband and tunable perfect optical absorber based on phase-change material," Nanoscale **12**, 5374–5379 (2020).

33. Y. Chen, X. Li, X. Luo, S. A. Maier, and M. Hong, "Tunable near-infrared plasmonic perfect absorber based on phase-change materials," Photon. Res., PRJ **3**, 54–57 (2015).

34. S. Fan, W. Suh, and J. D. Joannopoulos, "Temporal coupled-mode theory for the Fano resonance

in optical resonators," JOSA A **20**, 569–572 (2003).

35. P. Hu, J. Wang, Q. Jiang, J. Wang, L. Shi, D. Han, Z. Q. Zhang, C. T. Chan, and J. Zi, "Global phase diagram of bound states in the continuum," Optica **9**, (2022).

36. M. Delaney, I. Zeimpekis, D. Lawson, D. W. Hewak, and O. L. Muskens, "A New Family of Ultralow Loss Reversible Phase-Change Materials for Photonic Integrated Circuits: $Sb_2S_3$ and $Sb_2Se_3$," Adv. Funct. Mater. **30**, 2002447 (2020).

37. J. P. Hugonin and P. Lalanne, *Reticolo Software for Grating Analysis* (Institut d'Optique, 2005).

38. E. D. Palik, *Handbook of Optical Constants of Solids* (Academic Press, New York, 1985).

39. T. Yin, J. Gu, G. Wang, C. Gu, B. Chen, X. Shen, and Y. Chen, "Realizing multi-level phase-change storage by monatomic antimony," Appl. Phys. Lett. **125**, 241902 (2024).

40. L. Jia, L. Cheng, and W. Zheng, "8-nm narrowband photodetection in diamonds," Opto-Electron. Sci. **2**, 230010–11 (2023).

41. X. Tan, H. Zhang, J. Li, H. Wan, Q. Guo, H. Zhu, H. Liu, and F. Yi, "Non-dispersive infrared multi-gas sensing via nanoantenna integrated narrowband detectors," Nat Commun **11**, 5245 (2020).

42. H. Zhu, Q. Li, C. Zheng, Y. Hong, Z. Xu, H. Wang, W. Shen, S. Kaur, P. Ghosh, and M. Qiu, "High-temperature infrared camouflage with efficient thermal management," Light Sci Appl **9**, 60 (2020).

43. M. Lee, G. Kim, Y. Jung, K. R. Pyun, J. Lee, B.-W. Kim, and S. H. Ko, "Photonic structures in radiative cooling," Light Sci Appl **12**, 134 (2023).

44. K. Sun, Y. Cai, L. Huang, and Z. Han, "Ultra-narrowband and rainbow-free mid-infrared thermal emitters enabled by a flat band design in distorted photonic lattices," Nat. Commun. **15**, 4019 (2024).